\documentclass[prd,aps,floats,twocolumn,nofootinbib,superscriptaddress]{revtex4}
\usepackage{commath}
\pdfoutput=1
\usepackage[usenames, dvipsnames]{color}
\usepackage{graphicx, array}
\usepackage{multirow}
\usepackage{graphics}
\usepackage{graphics,epsfig}
\usepackage{amssymb}
\usepackage{amsmath}
\usepackage{ulem}
\usepackage{multirow}
\usepackage{subfigure}
\usepackage{bm}
\usepackage{txfonts}
\usepackage{cancel}
\usepackage{url}
\usepackage{hyperref}
\usepackage{lipsum}
\usepackage{graphicx}
\usepackage{comment} 
\usepackage{soul}

\newcolumntype{C}[1]{>{\centering\arraybackslash}m{#1}}

\def\be{\begin{equation}}
\def\ee{\end{equation}}
\def\bi{\begin{itemize}}
\def\ei{\end{itemize}}
\def\ben{\begin{enumerate}}
\def\een{\end{enumerate}}
\def\bt{\begin{tabular}}
\def\et{\end{tabular}}
\def\bc{\begin{center}}
\def\ec{\end{center}}

\def\bea{\begin{eqnarray}}
\def\eea{\end{eqnarray}}

\def\ba{\begin{eqnarray}}
\def\ea{\end{eqnarray}}

\let\oldhat\hat

\renewcommand{\hat}[1]{\oldhat{\boldsymbol{\mathbf{#1}}}}

\graphicspath{ {./images/},{ ./images/ratioFigs/},{ ./images/windowFigs/},{ ./Version1/}}

\begin{document}
\input{epsf}

\title{Cluster optical depth and pairwise velocity estimation using machine learning}

\author{Yulin Gong}
\author{Rachel Bean}
\affiliation{Department of Astronomy, Cornell University, Ithaca, NY 14853, USA}

\begin{abstract} 
We apply two machine learning methods, a CNN deep-learning model and a gradient-boosting decision tree, to estimate individual cluster optical depths from observed properties derived from multiple complementary datasets. The models are trained and tested with simulated N-body derived halo catalogs and synthetic full-sky CMB maps designed to mirror data from the DESI and Simons Observatory experiments. Specifically, the thermal Sunyaev-Zel'dovich (tSZ) and CMB lensing convergence, along with cluster virial mass estimates are used as features to train the machine learning models. The predicted optical depths are combined with kinematic Sunyaev-Zel'dovich (kSZ) measurements to estimate individual cluster radial peculiar velocities. The method is shown to recover an unbiased estimate of the pairwise velocity statistics of the simulated cluster sample. The models are demonstrated to be effective for halos with mass range $10^{13} M_{\odot} < M_{200} < 10^{15} M_{\odot}$ over a redshift range $0<z<1$, and validated in the presence of primary CMB, instrument noise, lensing convergence noise, and potential uncertainties in halo virial mass estimates. We apply the method to ACT CMB data, using ACT DR4 component-separated maps for tSZ and CMB lensing and ACT DR5 maps for kSZ, in conjunction with galaxy clusters observed in the SDSS DR15 spectroscopic survey. We demonstrate that the machine learning approach is an effective one to analyze data from current and upcoming CMB experiments such as Simons Observatory and CCAT, and galaxy surveys, such as DESI and Roman, for which the pairwise velocity statistics can provide valuable insights into the properties of neutrinos and gravity on cosmic scales.
\end{abstract}. 

\maketitle
\section{Introduction}
\label{sec:intro}
Measurements of the cosmic microwave background (CMB) \cite{WMAP:2003ivt, WMAP:2010qai, WMAP:2012nax, Planck:2013pxb, Planck:2015mrs, Planck:2018vyg, AtacamaCosmologyTelescope:2013swu, ACT:2020gnv, ACT:2023kun,ACT:2025fju,ACT:2025tim}, baryon acoustic oscillations \cite{SDSS:2005xqv, Percival:2007yw, SDSS:2009ocz, Addison:2013haa, BOSS:2014hhw, Kazin:2014qga, Cuceu:2019for, eBOSS:2020yzd, DESI:2024mwx}, and Type Ia supernovae \cite{SupernovaSearchTeam:1998fmf, SupernovaCosmologyProject:1998vns, Zhang:2017aqn, Brout:2022vxf, DES:2022tjd, DES:2024tys, DES:2024hip} have provided a multi-prong approach to understanding the matter constituents of the universe and cosmological evolution, including the current accelerative expansion \citep{Peebles:2002gy, Copeland:2006wr, Frieman:2008sn, Weinberg:2013agg}. $\Lambda$CDM, comprising Cold Dark Matter (CDM) and a Cosmological Constant, $\Lambda$, as well as Standard Model matter, is the prevailing cosmological model, consistent with the observations. Modifications to General Relativity (GR) have also been considered, as alternatives to $\Lambda$, to explain the cosmic acceleration e.g. \citep{Clifton:2011jh, Joyce:2014kja, Joyce:2016vqv, Nojiri:2017ncd, Langlois:2018dxi}. Measurements of the clustering of galaxies and galaxy clusters on cosmological scales provide powerful tools to compare and contrast modified gravity theories and the standard cosmological model, allowing for the examination of the growth of cosmic structures over time \citep{Bellini:2014fua, Bull:2015stt, deMartino:2015zsa, Lombriser:2016yzn, Amendola:2017orw, Cusin:2017wjg, Perenon:2019dpc, Perenon:2021uom, Jiang:2023nzz}. 

Through their mutual gravitational attraction, galaxy clusters move towards each other inducing pairwise statistics \citep{Ferreira:1998id, Diaferio:1999ig, Aghanim:2001yu}. The pairwise peculiar cluster velocities can provide insights into the dynamics of large-scale structure formation, gravity models, and the sum of neutrino masses \cite{Bhattacharya:2006ke, Bhattacharya:2007sk, Keisler:2012eg, Mueller:2014dba, Mueller:2014nsa, Alonso:2016jpy, Kuruvilla:2020gcm, Zheng:2020qcw}. The measurement of such peculiar velocities of galaxy clusters is challenging, however, and relies on indirect inference from cluster observables. 

The Sunyaev-Zeldovich (SZ) effect \citep{Sunyaev:1970eu, Sunyaev:1972eq, Sunyaev:1980nv} is a key cluster observable resulting from the CMB photons interacting with hot electrons in the intra-cluster medium (ICM). The SZ effect can be divided into two main forms: the thermal SZ effect (tSZ) and the kinetic SZ effect (kSZ). Both effects create a CMB temperature shift where the tSZ effect is caused by the inverse Compton scattering of CMB photons with high-energy electrons in galaxy clusters, and the kSZ effect is a Doppler shift caused by the bulk motion of ICM within galaxy clusters relative to the homogeneous rest frame. The kSZ effect therefore provides a natural probe for the peculiar velocities of clusters along the line of sight. 

A pairwise kSZ momentum estimator has been widely utilized to detect the kSZ signal. A detection of the kSZ effect with the pairwise momentum statistic was first measured in \citep{Hand:2012ui} from Atacama Cosmology Telescope (ACT) CMB data \citep{Swetz:2010fy} with the Sloan Digital Sky Survey (SDSS) data \citep{SDSS:2006srq}. More recent measurements of the pairwise kSZ with ACT and SDSS data have also been made  \citep{DeBernardis:2016pdv, Calafut:2021wkx} and studied in tandem with measurements of the tSZ \cite{DeBernardis:2016pdv, Vavagiakis:2021ilq, Liu:2025zqo}. The detection of the kSZ signal has also been reported with Planck CMB and SDSS galaxy data \cite{Planck:2015ywj}, South Pole Telescope (SPT) \cite{Carlstrom:2009um} and the Dark Energy Survey (DES) cluster data  \citep{DES:2016umt, SPT-3G:2022zrq} and Planck and the Dark Energy Spectroscopic Instrument (DESI)\footnote{\url{https://www.desi.lbl.gov/}}\citep{DESICollaboration2016} Legacy Imaging Survey \cite{Li:2024svf}. 

Other measurements using the kSZ include tomography \citep{Smith:2018bpn, Munchmeyer:2018eey, Contreras:2019bxy, Pan:2019dax, Chaves-Montero:2019isa, Sato-Polito:2020cil, Kumar:2022bly, AnilKumar:2022flx, Lee:2022udm}, projected fields \citep{Hill:2016dta, Ferraro:2016ymw, Kusiak:2021hai, LaPlante:2021ced, Bolliet:2022pze}, velocity reconstruction \citep{Li:2014mja, ACTPol:2015teu, Tanimura:2022fde, Nguyen:2020yuc, DES:2023mug, Gong:2024elx}, kSZ detections for individual clusters \citep{Sayers:2013ona}, stacked clusters \citep{AtacamaCosmologyTelescope:2020wtv}, power spectrum measurements \citep{George:2014oba}, kSZ cluster velocity dispersion \citep{Planck:2017xaj}, and 21cm-kSZ correlations \citep{Li:2018izh}. 

Commonly, the pairwise kSZ momentum estimator is combined with optical depth estimates,  derived for example from the tSZ and X-ray observations, to obtain a pairwise velocity estimator. Specifically, the tSZ Compton y-$\tau$ scaling relation \cite{Battaglia:2016xbi, Hadzhiyska:2023cjj} is one of the most commonly used methods to infer the mean sample-averaged optical depth of galaxy clusters from tSZ observations. 

In general, scaling relations are simple empirical relationships observed between different physical properties of interest. Without necessarily comprehending all of the underlying physics involved, these relations enable the inference of one property of an object based on the measurement of another. Examples of cluster scaling relations include the mass-luminosity relation \citep{Bell:2000jt, Bell:2003cj, Chabrier:2003ki, Bernardi_2010}, stellar to halo mass \cite{Kravtsov:2014sra, Wechsler:2018pic, Girelli:2020goz}, Compton y-mass  \citep{Planck:2012zvb, Battaglia:2014cga, Pop:2022mfn}, and mass-richness \citep{DES:2018kma, Murata:2019fxk, Abdullah:2022eqa}.

In physics, machine learning (ML) has been widely used to uncover patterns and insights in large and complex datasets that would be computationally impossible to obtain through other means e.g. \cite{10.3389/fphy.2024.1322162}. In CMB analysis, this has included applications to clean foreground contamination \citep{Wang:2022ybb}, component separation \cite{Casas:2022teu} and CMB delensing \cite{Ni:2023ume}. Machine learning methods have been used to recover cluster masses \citep{Ho:2019zap, Green:2019uup, Yan:2020wsr, KodiRamanah:2020tlp, Ho:2020lzz, deAndres:2021tjl, Ho:2023dgt,Zhao:2024eaj} including from tSZ signals \citep{Gupta:2020yvd, deAndres:2022mox, Wadekar:2022cyw}. ML has also been used in the context of kSZ: estimating galaxy cluster peculiar velocities from the surrounding galaxy distribution \cite{Tanimura:2022fde}, and modeling individual isolated clusters \cite{Wang:2020kvd, Wadekar:2022cyw}.

Here we extend on this prior work by evaluating the effectiveness of machine learning to infer cluster properties from realistic synthetic projected CMB maps of tSZ and CMB gravitational lensing. The intent is to apply this to the analysis of current and upcoming CMB surveys such as the Simons Observatory (SO)\footnote{\url{https:/simonsobservatory.org/}}\citep{SimonsObservatory:2018koc}. In comparison to previous work that focused on machine learning models trained using 3D and/or analytical cluster information for clusters treated in isolation from another, our approach uses 2D maps, projected along the line of sight, to mirror CMB survey data. The projected signal includes the potential for contamination from overlapping clusters along the line of sight \citep{Gong:2023hse} and the two-halo term \citep{Vikram:2016dpo, Hill:2017tua, Amodeo:2020mmu}. We also include primary CMB and instrument noise to mimic expected survey properties for SO. We analyze galaxy clusters with a lower mass threshold than in previous analyses, extending to $\sim 10^{13}M_{\odot}$, and over a redshift range $0<z<1$ to model datasets, such as those from DESI.

In our analysis, we consider two distinct machine learning models. Convolution Neural Network (CNN) is one of the most widely used machine learning models and was used in previous studies of cluster mass estimation, as cited above, primarily utilizing the VGG (Visual Geometry Group) architecture \citep{simonyan2015deep}, a deep convolutional neural network known for its simplicity and uniform use of small filters, achieving strong performance in image recognition tasks. CNN-VGG models, however, have several limitations and problems that hinder performance enhancement, necessitating the exploration of more sophisticated CNN architectures \citep{deAndres:2022mox}. In this work, we employ a more advanced CNN architecture, ResNet (Residual Network)\citep{7780459}, which is designed to address the issues inherent in VGG models. We also utilize the Gradient Boosting Decision Trees (GBDT) technique, a ML method widely applied in astronomy (e.g., \citep{Li_2021, Sahakyan:2022uvr, Tolamatti:2023hef, Darya_2023, Coronado-Blazquez:2023bbu, Gong:2024elx, Euclid:2024fdu}). The rationale for this additional ML model is that CNNs can perform poorly when images are contaminated by extraneous signals, such as the primordial CMB and detector noise \citep{yim2017enhancingperformanceconvolutionalneural,hosseini2017limitationconvolutionalneuralnetworks,Su_2019}. 

We compare and discuss the performance of the two machine learning models in the context of cluster mass and optical depth inference. We demonstrate the feasibility of extracting cluster mass from tSZ measurements, and cluster optical depth from tSZ combined with CMB lensing convergence and cluster virial masses using the CNN-ResNet and GBDT architectures.  We consider how the machine learning optical depth estimates can be combined with the kSZ signatures to infer the peculiar velocities of galaxy clusters, and the pairwise velocity estimator. We test the model effectiveness when systematic noise such as the primary CMB, instrument noise, and lensing convergence noise are present, as well as other systematic effects such as scattering for the cluster virial mass estimates. In addition, we introduce a novel technique using an Autoencoder machine learning model to deproject the line-of-sight contributions to the signal from high-redshift structures.

The structure of this paper is as follows: Section \ref{sec:formalism} discusses the formalism related to our study and Section \ref{sec:data} describes the datasets used. The machine learning models are detailed in Section \ref{sec:ML} and the results are presented in Section \ref{sec:results}. We summarize the conclusions of our study in Section \ref{sec:conc}.

\section{Formalism}
\label{sec:formalism}

\subsection{CMB Observables}
\label{sec:SZ}
The kSZ and tSZ effects are related to the interaction between the CMB radiation and, respectively, the momentum and pressure of the moving electrons in matter along the line of sight. The kSZ can be written as
\begin{equation}
\label{eq:kSZ}
    b(\hat{n})= -\frac{\delta T_{kSZ}}{T_{CMB}}(\hat{n}) = - \frac{\sigma_{T}}{c}\int \frac{d\chi}{1+z} e^{-\tau(z)} n_e(\chi\hat{n},z)\mathbf{v_e}(\chi\hat{n},z)\cdot\hat{n},
\end{equation}
where c is the speed of light, $\sigma_T$ is the Thomson cross section, and $T_{CMB}$ is the average temperature of the CMB anisotropy. The CMB signal in sky direction, $\hat{n}$, is a line-of-sight integral over matter at a comoving radial distance, $\chi(z)$, at redshift $z$. The number density of electrons and peculiar velocity, of the matter are $n_e$ and $v_e$, respectively. The optical depth, $\tau$, for the matter can then be defined as
\begin{equation}
\label{eq:tau}
\tau (\hat{n}) = \sigma_T \int \frac{d\chi}{1+z} n_e(\hat{n}\chi,z).
\end{equation}
Clusters are optically thin, $\tau\sim 10^{-3}$, so that it is reasonable to assume $e^{-\tau}\approx 1$ to sub-percent level accuracy.

The Compton $y$-parameter from the tSZ effect can be similarly expressed as 
\begin{equation}
\label{eq:tSZ}
  y (\hat{n})= \frac{1}{f_{SZ}} \frac{\delta T_{tSZ}}{T_{CMB}}(\hat{n})=\frac{\sigma_T}{m_e c^2}  \int \frac{d\chi}{1+z} P_{e}(\chi\hat{n},z),
\end{equation}
where $m_e$ is the electron rest mass, $P_e$ is the electron pressure and $f_{SZ}$ is a frequency dependent factor,
\begin{equation}
f_{SZ} = x\frac{e^{x}+1}{e^x-1}-4
\end{equation}
with $x = h\nu/k_BT_{CMB}$ for a frequency $\nu$.

We also include a third secondary CMB anisotropy, CMB gravitational lensing from the deflection of CMB photons by large-scale structure of matter in the path. The CMB lensing convergence parameter, $\kappa$, can be used to characterize this and is determined by the integrated mass along the line of sight,
\begin{equation}
\label{eq:kappa}
    \kappa(\hat{n}) = \frac{3\Omega_m H_0^2}{2c^2} \int_0^{\chi_{*}} d\chi (1+z) \frac{\chi(\chi_{*} - \chi)}{\chi_{*}} \delta(\chi \hat{n}, z),
\end{equation}
where $\Omega_m$ is the matter density, $H_0$ is the Hubble constant, $\chi_*$ is the comoving distance to the surface of the CMB last scattering, and $\delta$ is the matter over-density.

We consider the average signal within a disk of aperture $\theta$, $\bar{\kappa}_{disk(\theta)}$, $\bar{y}_{\theta}$ and $\bar{b}_{\theta}$, for two choices of angular radius: 1) an aperture of fixed angular extent, e.g. $\theta = 2.1'$, typically used in analyzing observations, 2) an aperture determined by each cluster's size, and varies across the sample, often used to analyze simulations, e.g. $\theta_{200} =  R_{200}/d_A(z)$, where $R_{200}$ is the physical radius within which the average cluster density is 200 times the critical density. For the kSZ and tSZ temperatures, we average pixel temperatures within angular apertures. 

The observed kSZ and tSZ signals are measured as quantities integrated within a cylinder of angular radius $\theta$,
\begin{eqnarray}
\label{eq:int_kSZ}
    B_{\theta}
    &=& -\frac{\sigma_T}{c}\int_{V_{cyl}} \frac{\chi^2 d\chi d\Omega}{(1+z)^3} n_e(\chi\hat{n},z) \mathbf{v_e}(\chi\hat{n},z)\cdot\hat{n},
\\
\label{eq:int_tSZ}
Y_{\theta} &=& \frac{\sigma_T}{m_e c^2}  \int_{V_{cyl}}\frac{\chi^2 d\chi d\Omega}{(1+z)^3} P_e(\chi\hat{n}\chi,z),
\end{eqnarray}
where $d\Omega$ is the solid angle. The cylindrical measures inherently include contributions from multiple halos falling within the aperture along the line of sight. Note cylindrical quantities are distinct from spherical quantities calculated by integrating radially outwards from the center of an individual cluster. 

In this work, we are interested in extracting cluster peculiar velocity information from the kSZ and tSZ measurements. In the limit where a single cluster at $z$ is the dominant contributor to the SZ effect along a given line of sight, the aperture-averaged kSZ and tSZ parameters can be estimated as $\bar{b}_{\theta}=B_{\theta}/(\pi \theta^2 d_A(z)^2) $ and $\bar{y}_{\theta}=Y_{\theta}/(\pi \theta^2 d_A(z)^2)$. The kSZ is then related to the cluster radial velocity, $v_r$, by
\begin{equation}
\label{eq:btauv}
   \bar{b}_{200} = \frac{v_{r}}{ c}\bar{\tau}_{200},
\end{equation}
where $\bar{\tau}_{200}$ is a velocity-weighted effective optical depth. 

Schaan et al. \cite{AtacamaCosmologyTelescope:2020wtv} applied such a velocity-weighted optical depth to stacked ACT SZ measurements of SDSS clusters, while Hadzhiyska et al. \citep{Hadzhiyska:2023cjj} showed that optical depth estimates from simulated stacked kSZ estimates were consistent with other data, such as X-rays and fast radio bursts. 

We use aperture photometry (AP) to reduce background contamination, such as from the primary CMB, in extracting the kSZ and tSZ signals $\delta T$, as defined in (\ref{eq:kSZ}) and (\ref{eq:tSZ}). This technique averages the signal coming from a circular disk of radius $\theta$ centered on the object of interest and then subtracts the average from an annulus region outside the disk:
\begin{equation}
\label{eq:AP}
    \delta \bar{T}_{AP(\theta)}(\hat{n}) = \delta \bar{T}_{disk(\theta)}(\hat{n}) - \delta \bar{T}_{annulus(\theta)}(\hat{n}).
\end{equation}
Typically, this annulus is directly adjacent to the disk and of equal area, with radius $\sqrt{2}\theta$. We will also discuss and compare other choices for ring annulus sizes in \ref{sec:training}. The AP-derived $\delta \bar{T}_{kSZ,AP(\theta)}$ and $\delta \bar{T}_{tSZ,AP(\theta)}$ can then be used, as in (\ref{eq:kSZ}) and (\ref{eq:tSZ}), to estimate $\bar{b}_{AP(\theta)}$ and $\bar{y}_{AP(\theta)}$.

\subsection{Pairwise velocity}
\label{sec:pairwise}
The pairwise velocity estimator \citep{Ferreira:1998id} is given by
\begin{equation}\label{eq:Vhat}
    {V} (r)= -\frac{\sum_{ij}(v_{i} - v_{j})c_{ij}}{\sum_{ij}c_{ij}^2},
\end{equation}
where $v_i$ is the line-of-sight peculiar velocity of a cluster i, $r$ is the distance between the cluster pair, $r \hat{n}_{ij}=\chi_i\hat{n}_i-\chi_j\hat{n}_j$, and
$c_{ij}$ accounts for the line-of-sight projection
\begin{equation}
    c_{ij}=\hat{n}_{ij}.\frac{\hat{n}_i-\hat{n}_j}{2}=\frac{(\chi_i-\chi_j)(1+\cos \ \alpha)}{2\sqrt{\chi_i^2\,+\chi_j^2\,-2 \chi_i \chi_j \cos \alpha}},
\end{equation}
with $\cos\alpha=\hat{n}_i.\hat{n}_j$.

Following \citep{Calafut:2021wkx}, we estimate the pairwise velocity in seventeen bins of pair separation: thirteen bins from $r$=20 to 150Mpc with a 10Mpc equal bin width and four unequal width bins with edges [150, 200, 250, 315, 395]Mpc.

$V$ is commonly inferred from the pairwise kSZ momentum estimator \citep{Hand:2012ui},
\begin{equation}\label{eq:pkSZ}
    P_{kSZ} (r,z_i)= -\frac{\sum_{ij}(\delta \bar{T}_{kSZ,i} - \delta \bar{T}_{kSZ,j})c_{ij}}{\sum_{ij}c_{ij}^2},
\end{equation}
which can be combined with an effective mass-averaged optical depth for the same cluster sample, $\bar{\tau}_{eff}$, to give
\begin{equation} \label{eq:PVtau}
    V(r) = -  \frac{c}{T_0}\frac{P_{kSZ}(r)}{\bar{\tau}_{eff}} .
\end{equation} 

Alternatively, in this work,  we train ML algorithms to predict velocity-weighted optical depths, $\bar{\tau}_{200}$, for each individual galaxy cluster using measurements of their $\bar{y}_{AP(\theta)}$, $\bar{\kappa}_{disk (\theta)}$ and estimated mass. The optical depth predictions are combined with kSZ measurements, $\bar{b}_{AP(\theta)}$, to infer peculiar radial velocities, as in (\ref{eq:btauv}). These velocities are then directly used in (\ref{eq:Vhat}) to estimate $V$. We determine how well ${V}_{pred}$, using the ML predicted velocities, $\{v_{pred}\}$, can recover an unbiased estimate of the true value, $V_{true}$ using the true halo velocities, $\{v_{true}\}$.

When analyzing real CMB observations, we compare ${V}_{pred}$ against a theoretical prediction, ${V}_{theory}$, from linear theory \citep{Sheth:2000ff} using the best fit Planck
cosmological model \cite{Planck:2018vyg}. Following \citep{Mueller:2014dba,Mueller:2014nsa}, 
\begin{equation}
{V}_{theory}(r,z) = -\frac{2}{3} \, f(z) \, H(z)  \,\frac{r}{1+z} \frac{\bar{\xi}_h(r,z)}{1 + \xi_h(r,z)},
\end{equation}
where $H(z)$ is the Hubble rate and $f(z)=d \ln\delta/d\ln a$ is the logarithmic growth rate. $\xi_h$
is the halo 2-point correlation function and $\bar{\xi}_h$ is averaged over the volume, defined as,
\begin{equation}
\xi(r, a) = \frac{1}{2\pi^2} \int dk \, k^2 j_0(kr) P_{lin}(k, z) b^{(2)}_h(k),
\end{equation}
\begin{equation}
\bar{\xi}(r, a) = \frac{3}{r^3} \int_{0}^{r} dr' \, r'^2 \xi(r', z) b^{(1)}_h(k).
\end{equation}
Here $P_{lin}$ is the linear matter power spectrum, $j_0$ is the spherical Bessel function, $b^{(m)}_h$ is the mass-averaged halo bias moments that is given by,
\begin{equation}
b_h^{(m)} = \frac{\int_{M_{\text{min}}}^{M_{\text{max}}} dM \, M \, n(M) \, b^{m}(M) W^2[kR(M)]}{\int_{M_{\text{min}}}^{M_{\text{max}}} dM \, M \, n(M) \, W^2[kR(M)]},
\end{equation}
where $n(M)$ is halo number density, $W(x)=3(\sin x - x\cos x)/x^3$ is the top-hat window function, and $R$ is the radial scale of the halo. 

\section{Data}
\label{sec:data}

\subsection{Simulated maps}
We use the publicly available full sky SZ simulation maps and cluster data from the Simons Observatory (SO) forecasts \citep{Sehgal_2010, SimonsObservatory:2018koc}. The simulation covers a volume of (1Gpc $h^{-1}$)$^3$ using 1024$^3$ dark matter particles, with a cosmology consistent with the WMAP 5-year parameter measurements \citep{WMAP:2008lyn}. The gas prescription for the SZ simulation uses a hydrostatic equilibrium model  \citep{Ostriker:2005ff, Bode:2009gv}, under the assumption that the gas first traces the dark matter and then quickly reorganizes into the hydrostatic equilibrium. The derivation of the deflection field for the CMB lensing convergence follows \citep{Das:2007eu}.

We use a halo and mock galaxy catalog developed with a friend-of-friends (FOF) halo finder provided as part of the simulation suite. We consider a halo sample with a redshift range of $0<z<1$ and mass range of $10^{13} M_{\odot} < M_{200}  < 10^{15} M_{\odot}$, with mean mass $3\times10^{13} M_{\odot}$. Within the catalog, the simulation products include an estimate of the integrated kSZ and tSZ signals, defined in (\ref{eq:int_kSZ}) and (\ref{eq:int_tSZ}), for the individual halos within the virial radius, $R_{vir}$, and the radii within which the mean density is 200 and 500 times the critical density of the universe, $R_{200}$, and $R_{500}$. The halo virial mass and gas mass of the halos (within $R_{vir}$, $R_{200}$, and $R_{500}$) are also specified.

We simulate the telescope optical noise and CMB anisotropies for the kSZ and tSZ maps to test our model's performance on real CMB observations. The power spectrum of the CMB is generated using CAMB \citep{2011ascl.soft02026L} with the WMAP 5-year parameter for a $\Lambda$CDM universe model consistent with the  simulations. Anticipating an observation from Simons Observatory around 150 GHz, the optical noise is modeled as white noise at $6.3 \, \mu K$-arcmin, and a Gaussian beam with FWHM = 1.4$^\prime$ \citep{SimonsObservatory:2018koc}. We also simulate the lensing convergence noise using the lensing convergence noise spectra from Simons Observatory LAT component-separated maps that is publicly available\footnote{\url{https://github.com/simonsobs/so_noise_models}}\citep{SimonsObservatory:2018koc}. We use the Synfast function from Healpy \citep{2005ApJ...622..759G,Zonca2019} to generate the full-sky noise map from the spectra. We find that the variance of the $\kappa$ noise map derived from the SO spectra is comparable to the variance of the noise map derived from the lensing noise spectra from ACT\citep{Darwish:2020fwf}.

\subsection{CMB and galaxy survey data}

In addition to the simulated datasets, we analyze the ACT CMB data\footnote{\url{https://lambda.gsfc.nasa.gov/product/act/actpol_prod_table.html}}. We use DR4 component-separated maps \citep{Madhavacheril:2019nfz} that isolate the tSZ signal and CMB lensing convergence. The maps cover the BOSS North (BN) and D56 region of the sky with 2,089 square degrees. For the kSZ signal, we consider the co-added ACT DR5 maps \citep{Naess:2020wgi} that combine ACT and Planck at 150 GHz and cover the sky with about 21,100 square degrees including the BN region.

We analyze the cluster sample from the SDSS DR15 spectroscopic surveys \citep{Aguado_2019} overlapping with the BN and D56 region as used in \citep{Calafut:2021wkx, Vavagiakis:2021ilq}. This has a total of 117,384 galaxy clusters with redshift range $0.1 < z < 0.8$ and luminosity $L > 6.1 \times 10^{10} L_{\odot}$.

\section{Machine Learning Model}
\label{sec:ML}

In \ref{sec:train_test} we describe our training and validating process for the machine learning models in general. We introduce the two principle machine learning, CNN and GBDT, models used in this work in \ref{sec:GBDT} and \ref{sec:deep}, as well as the format of the input data used for each model. An Autoencoder model used to remove high-redshift contributions to the lensing convergence map is presented in \ref{sec:autoencoder}. We discuss the statistics used to determine model prediction performance and efficacy in \ref{sec:prediction_statistics}.

\subsection{Training and testing process}
\label{sec:train_test}
To train and test the ML models, we create a sample of 500,000 randomly selected halos from the SO forecast simulation catalog with $0 <z< 1$ and  mass threshold $M_{200} > 10^{13} M_{\odot}$. We split the sample into two for training and testing purposes, corresponding to a 70/30 percentage split. The training sample of 350,000 objects is used for the model to learn the data properties, while the testing sample (herein SO Test Sample), containing the remaining 150,000 objects, is used to study the trained model performance on unseen data.

Developing a model that can make precise predictions on unseen data is one of the main objectives of the training process. To achieve this, it is important to prevent over-fitting i.e. that while the model has good performance on training data it does not generalize well on new, unseen data. To address this, we train the model using the cross-validation technique. It involves partitioning the training data into multiple folds, training the model on some folds while evaluating its performance on others, and repeating this process several times. Specifically, we divide the training data into five folds with equal sample size, and in each iteration, we train the model using four folds while evaluating the model performance on the remaining one fold. Based on the average performance of these five iterations, we are able to tune the hyper-parameters of the ML model. Different from the model parameters, which are learned during training, hyper-parameters are not learned but are predefined and are initialized before learning begins. They can have a significant impact on the effectiveness and performance of the model and are essential in regulating how the learning algorithm behaves \citep{probst2018tunabilityimportancehyperparametersmachine, weerts2020importancetuninghyperparametersmachine, jin2022hyperparameterimportancemachinelearning}. We use the Python package OPTUNA\footnote{\url{https://optuna.org/}} to help select the model hyper-parameters that optimize the model's performance on unseen data. Once the hyper-parameters are determined, we train the model again now using these hyper-parameters on the full set of training samples to finalize our model. 
Another crucial component for the training process is the loss function, also known as the objective function. The loss function serves as a quantitative metric to minimize during training, which helps to direct the learning process by evaluating how well the model's predictions match the actual target values. For both the CNN and GBDT models, we use the mean square error (MSE) as our loss function, denoted as
\begin{equation}\label{eq:MSE}
    MSE = \frac{1}{N}\sum_{i=1}^{N}(x_{t,i} - {x}_{p,i})^2,
\end{equation}
where $N$ is the number of data points, $x_t$ is the true values, and ${x}_{p}$ is the predicted values. 

Here, the machine learning model learns from signals measured from the tSZ and CMB lensing convergence maps to predict the optical depth and halo mass for each galaxy cluster. For training, we use noise-free maps (e.g. pure tSZ maps without any statistical noise or systematic contamination) to allow the model to explicitly focus on learning the true relationships without any interference. While one might also choose to train on noisy data such as including primordial CMB and detector noise, we find that training on pure signal data consistently outperforms training on noisy data, regardless of whether noise is present in the test data or not. Training on pure signal has better performance because the model learns essential patterns without being misled by noise, leading to better generalization. In contrast, training on noisy data suffers from overfitting to noise-specific artifacts, which reduces the model's adaptability to both noisy and noise-free test conditions. For testing, we first test the model performance on signals measured from the noise-free maps. We then test with systematic effects included, such as instrument noise and CMB anisotropy residuals, to reflect the analysis one would expect from real observations with component-separated maps (e.g. ILC from ACT observations \citep{Madhavacheril:2019nfz}). In the next two subsections, in addition to introducing the ML models, we will also discuss in detail the format of the input data that is fed to the GBDT and deep learning CNN models respectively for both training and testing processes.

\subsection{Gradient boosting decision trees}
\label{sec:GBDT}
Gradient boosting decision trees (GBDT) is a powerful ML method that combines the concepts of decision trees with gradient boosting. To achieve prediction accuracy, it trains a sequence of decision trees one after the other, fixing mistakes committed by the previous tree. Through an emphasis on learning from errors, GBDT builds an ensemble model that is highly effective in regression.

The GBDT algorithm is implemented using a widely used Python package, LightGBM \citep{NIPS2017_6449f44a}\footnote{\url{https://lightgbm.readthedocs.io/en/stable/}}. The model hyper-parameters are obtained using the technique discussed in \ref{sec:train_test}. The model was trained using a learning rate of 0.01 and a maximum of 63 nodes per tree, which controls the complexity of the individual trees. To prevent overfitting, the minimum number of data points required in a leaf was set to 127. Additionally, early stopping was employed, with training stopped once the performance does not
improve for 50 iterations.  

\begin{figure*}
\includegraphics{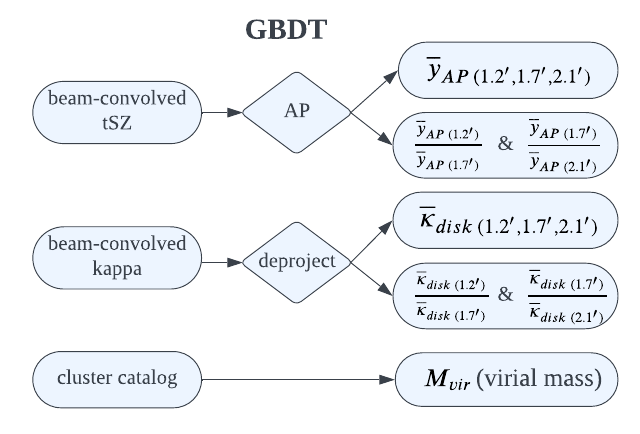}
\caption{The model features of our GBDT algorithm. From the tSZ map, we measure five signal features: the aperture photometry (AP) signals under three different aperture sizes (i.e. $1.2^\prime,1.7^\prime,2.1^\prime$) and two of the ratios between the three $\bar{y}_{ap}$. For the CMB lensing convergence map, we first deproject the map using Autoencoder and measure the average disk signal within the three aperture sizes. We also consider the ratio between the three $\bar{\kappa}_{disk}$. We also consider the halo virial mass as an additional feature.} 
\label{fig:FE_GBDT}
\end{figure*}

Feature engineering is a key process for the GBDT algorithm; it converts unprocessed input data into features that help the predictive models better understand the underlying problem and achieve better performance. The cluster features used for the GBDT model are summarized in Fig.~\ref{fig:FE_GBDT}.

The tSZ signal is a key feature for the GBDT model. It is measured from the component-separated tSZ map using aperture photometry under three different angular scales: $1.2^\prime, 1.7^\prime, 2.1^\prime$. The central value, $1.7'$, is equivalent to the average size of $\theta_{200}$ over the full sample, i.e. is an indicative measure of the cluster size. Including features computed at multiple scales provides richer, complementary information that helps generalize the model as well as enhancing the stability of the model prediction.  Following \citep{Wadekar:2022cyw}, we also consider the ratio between these three angular sizes to help the model capture multiscale information about the cluster concentration. The concentration parameter characterizes the distribution of mass and gas within the cluster, especially distinguishing between the dense core and the less dense outskirts. Concentration is useful as a feature because it encodes the internal distribution of gas, allowing predictive models to minimize uncertainties associated with variability in cluster cores, thus improving the accuracy and stability of model predictions. For the CMB lensing convergence map, we first use the Autoencoder model, to be discussed in \ref{sec:autoencoder}, to remove contributions to the $\kappa$ map from high redshifts. We then measure the average disk signal, $\overline{\kappa}_{disk}$, from the deprojected $\kappa$ map with the same three aperture sizes, as well as the ratio between the three, following the rationale for the tSZ features. Finally, we include the halo virial mass as an additional feature. Observationally, we assume the halo virial mass is being estimated from the optical and infrared luminosity and stellar mass estimates, e.g. \citep{Wen:2024gho,Kravtsov:2014sra}.

We employ the GBDT model for two applications. Our principal focus is the estimation of the optical depth, $\bar\tau_{200}$, using the full set of tSZ, lensing, and cluster mass features. We also, however, consider how the algorithm performs when trained on the tSZ data alone to estimate the cluster mass. We do this to benchmark the algorithm to see how well it can reconstruct known cluster properties from the catalog.

\subsection{Deep learning model - Convolutional Neural Network}
\label{sec:deep}
A neural network is a statistical technique in which computers mimic how the human brain biologically processes data and performs complex tasks. Specifically, a Convolutional Neural Network (CNN) is a type of neural network architecture that is designed for computer vision tasks. As discussed in \ref{sec:intro}, CNN is the most widely used ML method for cluster mass estimation using cluster images. Previous cluster mass estimation analyses used a VGG (Visual Geometry Group) CNN architecture \citep{simonyan2015deep}. The CNN-VGG model used here is built following \citep{deAndres:2022mox}.  We also use an alternative CNN architecture, Residual Networks (ResNet) \citep{7780459}, proposed to resolve issues present in a CNN-VGG model.  The ResNet model has more than 10 million trainable model parameters, while the VGG model has only 50,000, making the ResNet model 200 times more complex.  It takes roughly 50 times longer to train the ResNet model for one epoch than the VGG model; we train the ResNet model for 15 epochs and the VGG model for 200 epochs. 

There are two key issues with a VGG model: the vanishing gradients problem and the degradation problem. The vanishing gradients problem occurs when the gradients of the loss function become very small as they propagate through the network. This can lead to extremely slow convergence or even stagnation of the learning process in the earlier stages of the network. The degradation problem refers to when deeper networks perform worse than simple shallower ones, i.e. the performance of a deep neural network decreases as additional layers are added. This is counter-intuitive as, in theory, deeper networks ought to function better as they can potentially represent more complicated functions. By introducing shortcut connections that skip one or more layers, the ResNet architecture introduces the idea of a {\it residual learning unit} to resolve these issues. It facilitates the maintenance of the gradient flow through the network, enabling more efficient propagation, even in extremely deep networks. 

\begin{figure*}
\includegraphics{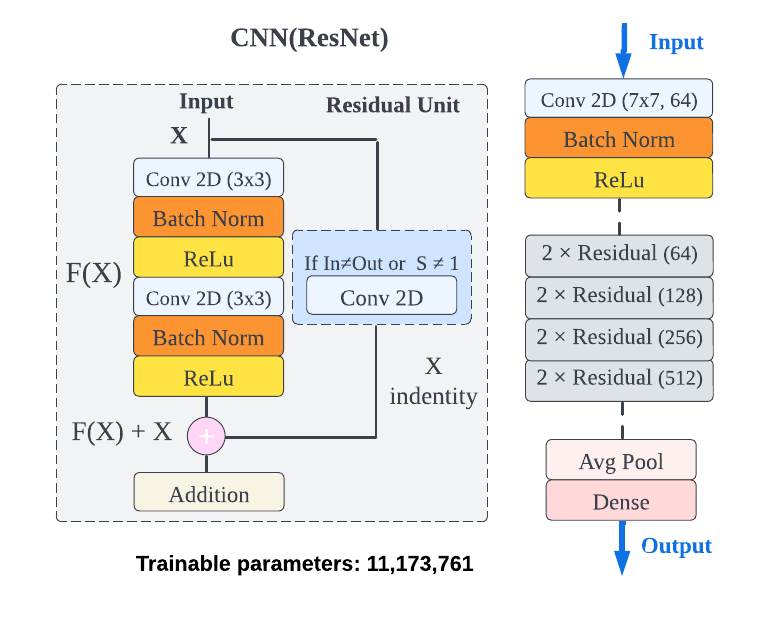}
\caption{The architecture of our CNN ResNet model. [Right] The model's overall workflow. [Left] The framework for the key component of a ResNet architecture, the residual unit. The total number of trainable parameters of the model are also summarized. The terms used in the figure are described in section \ref{sec:deep}. The number and size of filters for each layer are also listed in the figure.}
\label{fig:FE_CNN}
\end{figure*}

The architecture of our full CNN-ResNet model is presented in Fig.~\ref{fig:FE_CNN}. $Conv 2D$ refers to a convolution layer that utilizes a series of filters to extract features from the input data. The $Batch \, Norm$ layer helps to normalize the inputs of each layer, which accelerates and stabilizes the learning process. A $ReLU$ (Rectified Linear Unit) layer is an activation layer that adds nonlinearity to the model, allowing it to learn intricate patterns. Four stages of residual units with an increasing number of $Conv 2D$ filters (64, 128, 256, 512) are then used to progressively capture features from the image. Increasing the number of filters progressively from 64 to 512 allows the network to progressively extract and refine features of varying complexity, starting with simple patterns at an early stage and moving to more complex structures as we move deeper. The key feature of a residual unit, shown on the left of Fig.~\ref{fig:FE_CNN}, is the skip connection where the original input (i.e. X) is directly added to the output of the layers. The network only learns the residual mapping F(X) instead of the direct mapping from the input X to the desired output Z, where the residual mapping refers to the difference between the desired output and the input. For a variable X, representing the input to the residual unit, if Z is the initial desired output, the network fits the residual, F(X) = Z - X, which is then recast into Z = F(X) + X. In this way, the vanishing gradient issue is solved as the gradients can flow smoothly through the skip connections, allowing networks even with hundreds or thousands of layers to be trained.

Following the residual units, the final stage of the network is a $Avg \, Pool$ layer which is a type of pooling layer that reduces the spatial dimensions of data, which helps reduce the computational complexity and control overfitting. Specifically, it separates the input feature map into sub-regions and outputs the average value from each sub-region. This follows a $Dense$ layer, also known as a fully connected layer, that is used to aggregate features and generate the final output in the last stages of neural networks.  During training, the dataset is typically divided into smaller subsets called batches. Each batch is passed individually through the network, which helps mitigate the computational and memory costs associated with processing the entire dataset at once. We train the CNN-ResNet model using the default batch size of 32, and train the model for 50 epochs, where an epoch refers to a complete processing of all the small batches through the network once. We also employ the Adam optimizer in our model due to its robustness and efficiency in training deep learning networks. Adam (Adaptive Moment Estimation) \citep{kingma2017adammethodstochasticoptimization} integrates the benefits of both AdaGrad \citep{duchi2011adagrad} and RMSProp \citep{tieleman2012rmsprop} by adaptively adjusting the learning rates for each parameter using estimates of the first and second moments of the gradients. This approach facilitates faster convergence and enhances performance on complex learning tasks. 

Different from the GBDT model, we do not manually measure features for the CNN model. Instead, we directly feed the raw tSZ/CMB lensing convergence cutout images (16x16 pixels, or $\sim$14'x14'), generated around the central location of each cluster, into the model, and the convolution layer measures the features for the model to use. 

\subsection{Autoencoder for high-z $\kappa$ signal deprojection}
\label{sec:autoencoder}

The tSZ signal is dominated by the emission from localized hot gas in the galaxy clusters, which evolve significantly at late times/low redshifts. In contrast, CMB lensing is determined by the total dark matter distribution, not just the gas density, as shown in (\ref{eq:kappa}), and is integrated over a far broad redshift window function, from the surface of CMB last scattering to today. Projection effects of the large-scale structure from high redshifts ($z> 3$) contribute to the $\kappa$ signal present in the map \citep{Nakoneczny}, while we are interested solely in the lower redshift contributions from the clusters. While for the highest mass clusters the low redshift component can dominate, the high redshift contamination is particularly relevant for the galaxy groups and moderate mass clusters ($<10^{14}M_{\odot}$).

We develop an ML tool, an {\it autoencoder}, to filter (or `deproject') the high redshift contributions to the CMB lensing convergence signal to reveal those from lower redshifts. We consider two cluster samples to help us measure the $\kappa$ radial profile for training purposes. We first randomly select 10,000 halos with a mass range $10^{13} < M_{200} < 1.5 \cdot 10^{13} M_{\odot}$ to represent the group of halos with low masses, and then select a second sample of equal size with mass $M_{200} > 10^{14} M_{\odot}$ to represent massive halos. The average stacked radial profile from these two samples are respectively considered as the lower and upper limit boundary of the possible $\kappa$ radial profile for a cluster sample with mass above $10^{13} M_{\odot}$.

\begin{figure}
\includegraphics[width = \columnwidth]{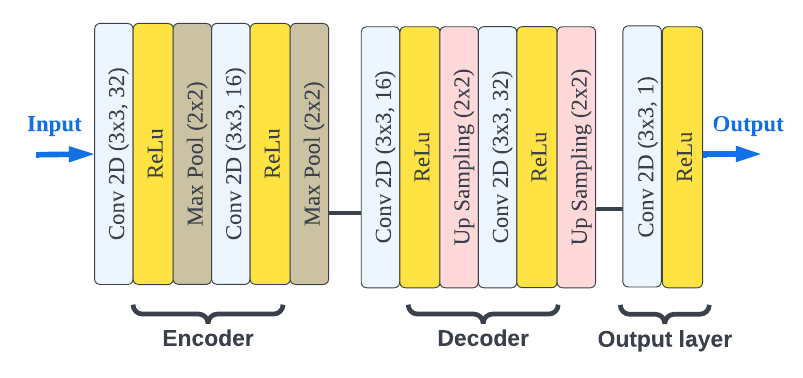}
\caption{The architecture and framework of our Autoencoder model. The model comprises three parts: the Encoder, Decoder, and the Output layer. The number and size of filters for each layer are also listed in the figure.}
\label{fig:autoencoder}
\end{figure}

An autoencoder is a type of artificial neural network that can be effectively used for denoising data, particularly images. The model is trained to reconstruct clean data from noisy inputs. The process is composed of encoding and decoding procedures and involves first transforming the data and then reconstructing the original data from the encoded representation. During the encoding process, the network compresses the input data into a compact representation of essential features. Then, in the decoding process, the compressed information is reconstructed back to recreate the clean data. We summarize our model in Fig.~\ref{fig:autoencoder}, where in addition to the layers described in \ref{sec:deep},the $Max \, Pool$ layer is another type of pooling layer that outputs the maximum value of each sub-region, and the $Up$ $Sampling$ layer is used to increase the spatial dimensions of the encoded compressed data. 

The training process is summarized below: 
\begin{enumerate}    
    \item We calculate the individual $\kappa$ profile as a function of radius (in arcmins) for each halo, and also the average radial profile over the 10,000 halos, for each of the two mass-selected cluster samples considered. In this, the contamination of the random deprojection effects sums to zero on average in the stacked $\kappa$ radial profile.

    \item We generate 100,000 random $\kappa$ radial profiles that are bounded by the two average $\kappa$ radial profiles obtained for the low and high-mass halos. These 1D profiles are used to create 2D images which we consider as the true $\kappa$ images without any line-of-sight contamination.
    
    \item We randomly select sky patches from the SO simulated $\kappa$ map, and add them on top of our halo-derived $\kappa$ images to mimic the additional line-of-sight contributions.
    
    \item Finally, the $\kappa$ images with line-of-sight contributions included are used as input and the true $\kappa$ images are used as output to train the autoencoder model. 
\end{enumerate}

\begin{figure}[!t]
\includegraphics[width = \columnwidth]{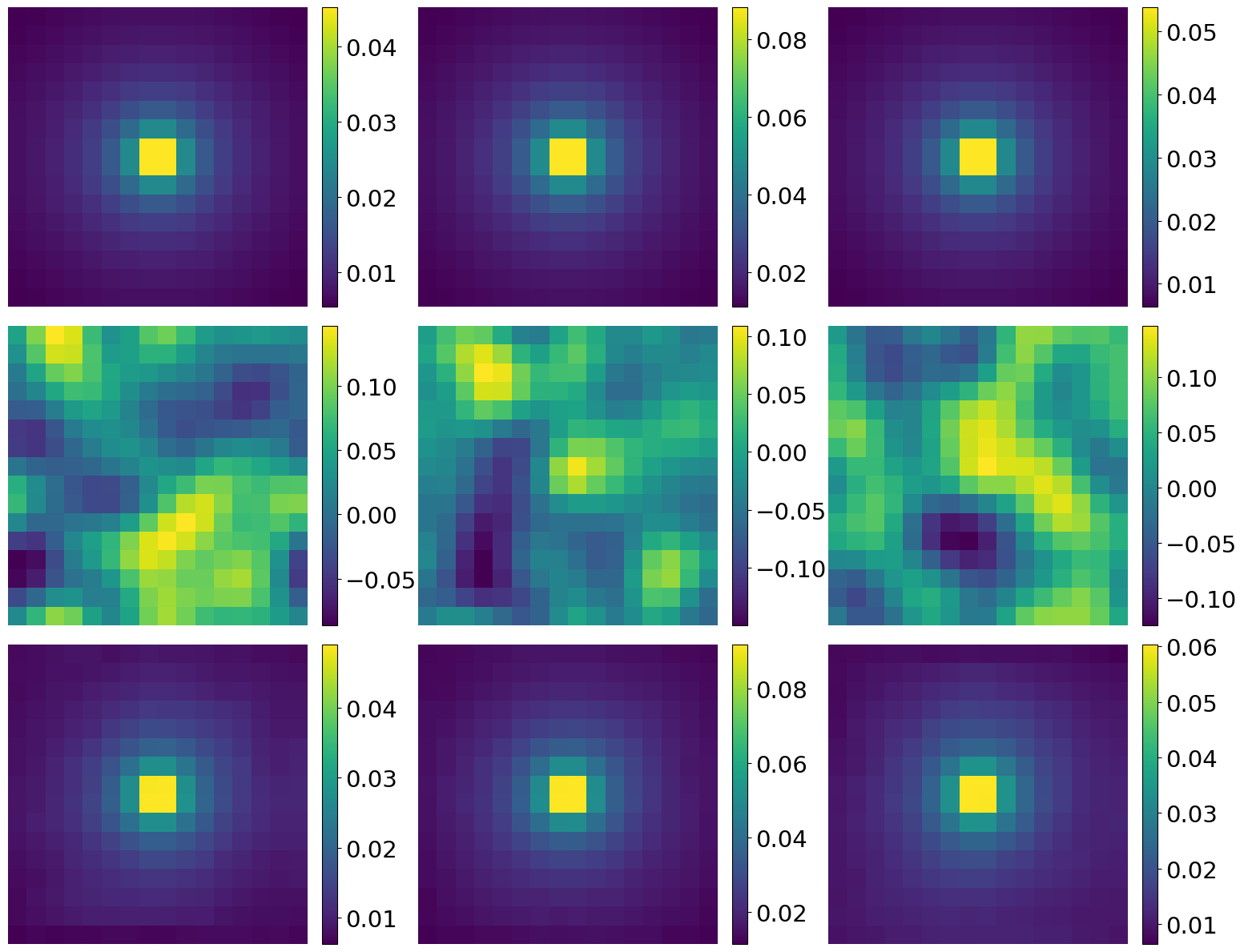}
\caption{ [Left to right] Three 16x16 pixel ($\sim 14x14'$) cutout images around target halos showing: [upper panel] the original CMB lensing convergence $\kappa$ map, [center] the original CMB lensing convergence $\kappa$ map with line of sight contributions added, and [lower] the corresponding deprojected images using the autoencoder model.}
\label{fig:denoise}
\end{figure}

We train the model for 30 epochs with a batch size of 32. The Adam optimizer is also used as discussed in \ref{sec:deep}. In Figure \ref{fig:denoise}, we show a comparison of the cutout images showing the true $\kappa$ signal, the signal after the line-of-sight contributions are added, and the signal recovered by the autoencoder for three example halos. 

\begin{figure*}
\includegraphics[width = 1.0\textwidth]{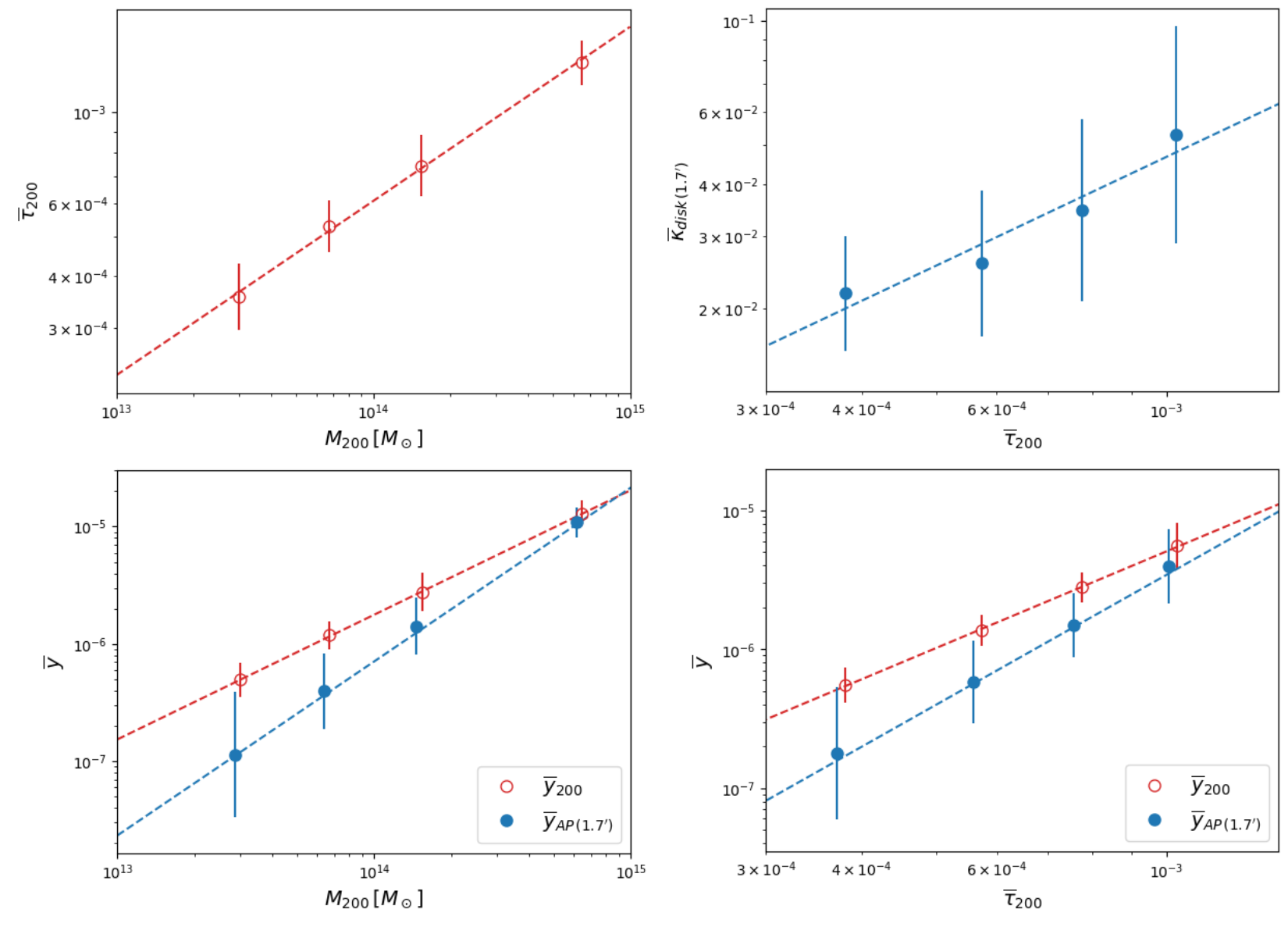}
\caption{Intrinsic relationships in the training data, showing the mean and 1$\sigma$ error bars for each bin, between the quantities the model will predict $\{M_{200}, \bar{\tau}_{200}\}$ and the training variables $\{ \bar{y}_{200}, \bar{y}_{AP}, \kappa_{disk}\}$ (both catalog-derived [red open marker] and map-derived [blue filled marker]).  [Upper left] The relationship between the cluster mass, $M_{200}$, and the velocity-averaged optical depth $\bar{\tau}_{200}$, and [upper right] $\bar{\tau}_{200}$ and the map-derived deprojected CMB lensing convergence, $\bar{\kappa}_{disk}(1.7')$ are shown.  The relationship between [lower left] $M_{200}$ and [lower right] $\bar\tau_{200}$ and the true tSZ y-parameter from the catalog, $\bar{y}_{200}$, and the map-derived y-parameter from aperture photometry, $\bar{y}_{AP (1.7')}$.}
\label{fig:tau_M}
\end{figure*}

\subsection{Statistics for model performance }
\label{sec:prediction_statistics}
We use Pearson's correlation coefficient between predicted values, $x_p$, and the true target values, $x$, obtained from the covariances for and between the respective datasets,
\begin{equation} \label{eq:correlation}
  corr (x,x_{p})=  \frac{cov(x,x_p)}{\sqrt{cov(x,x)\,cov(x_p,x_p)}}.
\end{equation}

We assess the extent to which the pairwise velocities predicted by our model, $V_{pred}$, agree with those directly derived from the simulation's true velocities, $V_{true}$, using the $\chi^2$,
\begin{equation} \label{eq:chi_v}
    \chi^2 = \sum_{ij} \Delta {V}_i\,{C}_{ij}^{-1}\,\Delta{V}_j,
\end{equation}
with $\Delta {V}_i = {V}_{i,true} - {V}_{i,pred}$.
The pairwise velocity covariance matrix is obtained via a bootstrapping technique by repeatedly drawing random samples (bootstrap samples) from the data with replacement. The covariance matrix ($C_{ij}$) is calculated from the lists of estimators for each of the bootstrap samples.

When analyzing real observational data, the covariance indicates the expected $1\sigma$ variation between the measurement and the theory, so that one expects a $\chi^2$ per degree of freedom of $\sim 1$ for an accurately estimated covariance and well-fitting theory. As noted in \citep{Gong:2024elx}, the $\chi^2$ here is used to compare the degree of agreement between the true and recovered statistic (which is determined by the performance of the ML algorithm) relative to the data covariance (which is unrelated to the ML performance). As such, it is used as a measure of how the errors in the ML prediction compare with the observational uncertainties. The desired scenario is that the errors in the ML prediction are far smaller than the measurement uncertainties, so that $\chi^2<1$, and in the limit of perfect prediction $\chi^2=0$.

\section{Results}
\label{sec:results}
In \ref{sec:training}, we outline relevant scaling relationships between the training variables and the quantities predicted from the machine learning model. In \ref{sec:tau_Vhat}, we present the model performance for cluster mass and optical depth estimation using multiple observables, and discuss the results of the pairwise velocity estimation with the simulated test data. In \ref{sec:ACT}, we apply the method to the real ACT CMB observations and SDSS galaxy cluster catalog.

\subsection{Intrinsic relationships between cluster variables}
\label{sec:training}

Prior to implementing the machine learning algorithm, we consider some of the intrinsic relationships between the cluster properties and observables used in the machine learning training. While the linear relationships described below are not explicitly used in this analysis, they provide a high-level heuristic connection to understand the data.

For the GBDT model, as discussed in \ref{sec:GBDT}, we use aperture photometry to measure the SZ features used to estimate cluster mass and optical depth. The traditional AP method uses a ring annulus directly adjacent to the central disk. \citet{Hadzhiyska:2023cjj} used a ring annulus detached from the central disk, with an inner radius of the ring annulus four times that of the central disk. The benefit of this separated annulus is that it can reduce the signal attenuation from the central disk. 

To understand how varying the distance between the center and the inner ring of the annulus affects the model's performance, we consider four different ring annulus distances for the AP method, with the inner ring radius ranging from 1 to 4 times the $r_{disk}$ for a sample of clusters with mass $\sim10^{14} M_\odot$. While the degree of signal attenuation decreases, we also find that the SNR between $\bar{y}_{AP}$ and $M_{200}$ decreases as the ring annulus is positioned further from the center, with SNR = 4.3, 2.8, 2.5, and 2.2 for the four ring annulus distances respectively. We also find that the correlation between $\bar{y}_{AP}$ and $M_{200}$ decreases as the size of the inner radius of the ring annulus increases. We therefore use the original AP method, with the ring annulus directly connected to the central disk, to measure the model features.

\begin{figure}[!t]
\includegraphics[width = \columnwidth]{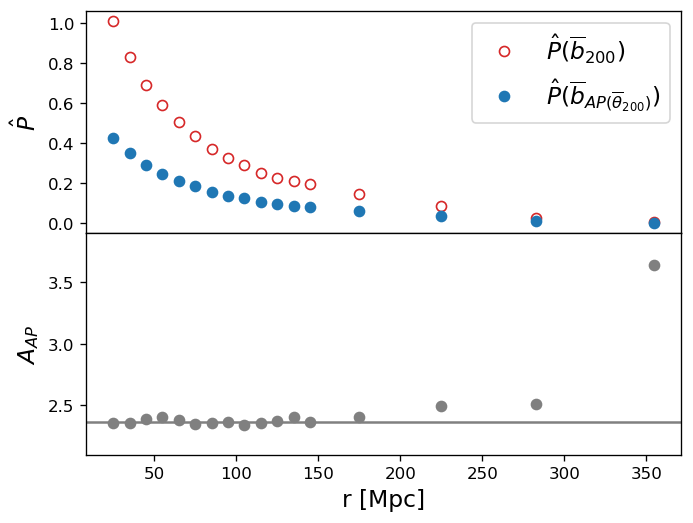}
\caption{[Upper] The pairwise momentum statistics as a function of cluster pair separation derived from $\bar{b}_{200}$ [red] and $\bar{b}_{AP(\overline{\theta}_{200})}$[blue] with $\bar{\theta}_{200}=1.7'$ for the training sample. [Lower] The attenuation factor, $A_{ap} = \hat{P}(\bar{b}_{200})/\hat{P}(\bar{b}_{AP(\bar{\theta}_{200})})$, and the constant value, $A_{ap}=2.37$, [gray line] fit to the data and used in the analysis.}
\label{fig:AP}
\end{figure}

In Fig.~\ref{fig:tau_M} we summarize some of the key scaling relationships between the simulation provided cluster mass, $M_{200}$, effective optical depth, $\bar\tau_{200}$, and tSZ amplitude, $\bar{y}_{200}$, and the map-based observables, specifically the tSZ y-parameter, $\bar{y}_{AP (1.7')}$ and the CMB lensing convergence, $\bar\kappa_{(1.7')}$. 

For cluster mass, $M_{200}$, we consider a linear regression relation, $a \cdot log(M_{200}) +b$, against the different variables:
\begin{eqnarray}
log(\bar{\tau}_{200}) &=& (0.42 \pm 0.01) log(M_{200}) - (9.14 \pm 0.19),
\\
log(\bar{y}_{200}) &=& (1.06 \pm 0.01) \, log(M_{200}) - (20.62 \pm 0.09),  \hspace{0.33cm}
\\
log(\bar{y}_{AP}) &=& (1.48 \pm 0.03) \, log(M_{200}) - (26.90 \pm 0.36),
\end{eqnarray}
with correlations of 0.87, 0.97 and 0.79 between the variable pairs, respectively. The cylindrical quantities, $\bar{y}_{200}$ and $\bar{\tau}_{200}$, have a strong correlation with the intrinsic cluster mass, $M_{200}$, from which we infer that cylindrical integration in itself isn’t a significant source of noise. 

The map-based observable, $\bar{y}_{AP}$, exhibits a correlation with $M_{200}$ that is 19$\%$ smaller than $y_{200}$. The AP signal for each individual halo includes contamination from co-aligned and nearby halos projected along the line of sight, and signal attenuation from the aperture photometry subtraction. The larger variance for the AP-measured signal leads to a lower correlation relation with cluster mass. 

We similarly fit linear regression relations for $\bar{\tau}_{200}$ and the other variables:
\begin{eqnarray}
    log(\bar{\kappa}_{disk}) &=& (0.88 \pm 0.18) \, log(\bar{\tau}_{200}) + (1.30 \pm 0.56),
    \\
    log(\bar{y}_{200}) &=& (2.32 \pm 0.04) \, log(\bar{\tau}_{200}) + (1.66 \pm 0.12), \hspace{0.25cm}
    \\
    log(\bar{y}_{AP}) &=& (3.12 \pm 0.09) \, log(\bar{\tau}_{200}) + (3.88 \pm 0.27),
    \end{eqnarray}
    with correlations 0.34, 0.83 and 0.56 between the variable pairs, respectively.
    
The correlation between $\bar{\tau}_{200}$ and $\bar{y}_{200}$  is 15\% lower than between $M_{200}$ and $\bar{y}_{200}$, which we attribute to the effect of line of sight contributions to the optical depth, unrelated to the target halo. The larger variance caused by the AP measurement, versus $\bar{y}_{200}$, together with the cylindrical integration reduce the correlation between $\bar{\tau}_{200}$ and $\bar{y}_{AP}$ to 0.56, 29$\%$ lower than the correlation between $M_{200}$ and $\bar{y}_{AP}$. Although the lensing convergence, $\kappa$, correlates comparatively weakly with optical depth, it nonetheless contains information and is included as a model feature for optical depth prediction. 

To derive velocity estimates, we combine the optical depth with kSZ measurements using AP with a fixed angular aperture, $\bar{b}_{AP(\theta)}$. Specifically, we use the average  $\theta_{200}$ across the test sample, $\theta = \bar{\theta}_{200} = 1.7'$. We compare the pairwise statistics derived using this fixed aperture versus the individual values of $\theta_{200}$ for each halo, and found the difference is less than $1\%$. We calculate and apply an attenuation factor to estimate the kSZ  signal in the disk from the kSZ AP signal which is attenuated through the ring annulus subtraction \citep{Gong:2023hse}. The attenuation factor is calculated as the ratio between the two pairwise statistics, i.e. $A_{ap} = \hat{P}(\bar{b}_{200})/\hat{P}(\bar{b}_{AP(\bar{\theta}_{200})})$,  derived from the two pure kSZ signals $\bar{b}_{200}$ and $\bar{b}_{AP(\bar{\theta}_{200})}$ of the training sample, without any noise or CMB signal contamination. Following \citep{Gong:2023hse}, we performed 1,000 bootstrap resamplings on $\bar{b}_{AP(1.7')}$ and computed $\hat{P}(\bar{b}_{AP(1.7')})$ for each bootstrap instant. We calculate $A_{AP}(r)$ for each bootstrap by comparing the bootstrap $\hat{P}$ to the $\hat{P}(\bar{b}_{200})$ calculated from the full training sample, and then determine the mean and standard error on $A_{AP}$. As shown in Fig.~\ref{fig:AP}, the attenuation factor calculated at each pair separation scale is found to be scale independent in the high signal to noise regime, for $r<250 Mpc$. We therefore treat $A_{ap}$ as a constant fit from the data, $A_{ap}=2.37\pm 0.01$.

\begin{figure*}[!t]
\includegraphics[width = \textwidth]{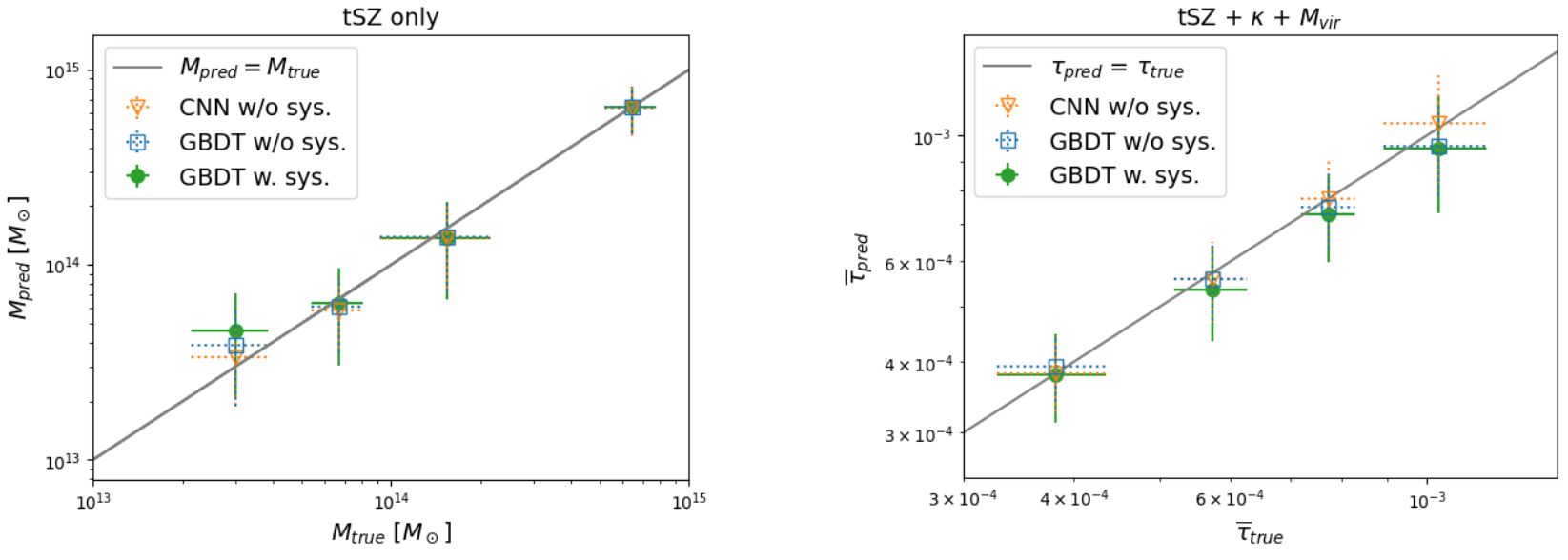}
\caption{[Left] Mean and 1$\sigma$ error bars for the predicted cluster mass, $M_{200}$, using only tSZ features, compared to the true mass for the SO Test Sample for the two different machine learning models. [Right] An analogous comparison but for the predicted velocity-weighted optical depth, $\bar{\tau}_{200}$, using tSZ, $\kappa$, and $M_{vir}$ features. Results are shown without systematic noise for CNN [orange triangle] and GBDT [blue square] and, for the GBDT,  with systematic uncertainties for the tSZ, lensing and virial mass [green circle].  }
\label{fig:tau_M_pred}
\end{figure*}

\begin{figure*}[!t]
\includegraphics[width = \textwidth]{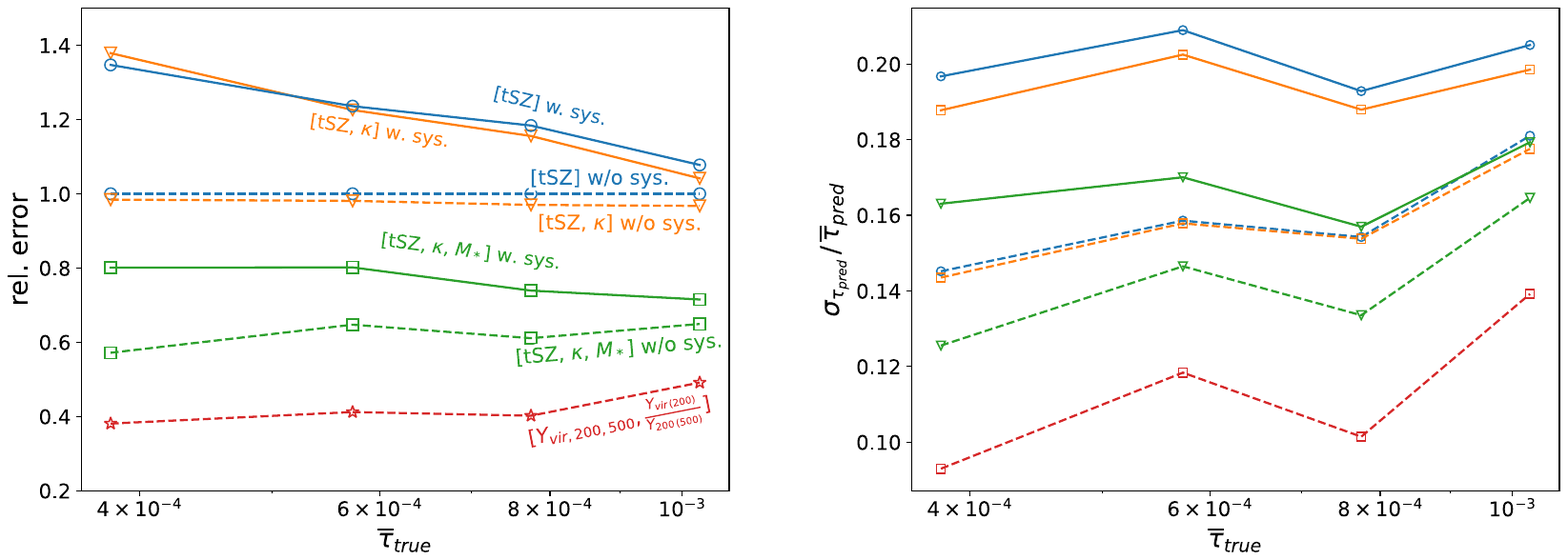}
\caption{[Left] Relative mean standard error (MSE) of cluster predicted velocity-weighted optical depth, $\bar{\tau}_{200}$, for the SO Test Sample using the GBDT model with different cluster feature sets as a function of the true optical depth. The prediction using the tSZ signal alone [blue dashed] is used as the baseline against which performance is compared for predictions from a model on tSZ and CMB lensing convergence, $\kappa$, [orange dashed], and trained on three observables: tSZ + CMB $\kappa$ + virial mass ($M_{vir}$) [green dashed]. Results are also shown with systemic uncertainties for tSZ, $\kappa$ and virial mass included [blue/orange/green solid]. An idealistic prediction if the spherical tSZ signals for the halos were available [red dashed]. [Right] The corresponding ratios between the 1$\sigma$ standard deviation and the mean optical depth estimate. } 
\label{fig:tau_feat_importance}
\end{figure*}

\subsection{Individual halo mass and optical depth estimation, and pairwise peculiar velocity reconstruction}
\label{sec:tau_Vhat}

We first consider the application of the machine learning algorithms to estimate the halo mass from the tSZ signal alone as a way to compare the performance of the CNN and GBDT architectures using a clearly defined cluster property that is strongly correlated with the input feature.

To test the performance of the two CNN architectures, we use 90,000 (30,000) randomly selected halos from the full 350,000 (150,000) training (testing) halo samples. This sample has a comparable sample size to the number of galaxy clusters from the SDSS surveys that is covered by the BOSS North region of the sky.

We find that the correlation between the predicted mass ($M_{pred}$) and the true mass ($M_{true}$) is 0.71 for CNN-VGG and 0.83 for CNN-ResNet model, corresponding to a 17$\%$ accuracy improvement using the CNN-ResNet model. Given this, in the rest of this work, we use the ResNet architecture for the implementation of CNN for the full training and testing samples. 

We train the CNN-ResNet and GBDT models using the full training set of 350,000 halos, training the CNN-ResNet model for 50 epochs and the GBDT model until the performance does not improve for 50 iterations, as outlined in \ref{sec:train_test}. 

In Fig.~\ref{fig:tau_M_pred}, we show the cluster mass estimation using the CNN-ResNet and GBDT models for the full SO Test Sample of 150,000 halos. We find that both models are able to efficiently provide an unbiased estimate of the cluster mass within 1$\sigma$ sampling uncertainty for the pure tSZ signal.  The correlation between the true mass and the predicted mass is comparable for the two models: 0.81 for GBDT and 0.86 for CNN-ResNet. The compute times to train the models, however, are significantly different, specifically the GBDT model is 9,000 times faster than the CNN-ResNet model. As such, the GBDT model has a clear pragmatic advantage.

We use the GBDT model to test the mass prediction performance when realistic noise is present for the tSZ. We assume component-separated tSZ maps are available (e.g. \citep{Planck:2015vgm, SPT-SZ:2021gsa, McCarthy:2024etq, ACT:2025xdm}) that include full detector noise and a possible 10$\%$ contribution of primary CMB residuals as systematic errors \citep{Madhavacheril:2019nfz}. We simulate one random realization of the detector noise and primary CMB, as described in \ref{sec:data}, to combine with the tSZ map.
The GBDT mass predictions from the data with systematic errors included are found to be consistent with the true values to well within the 1$\sigma$ variation in each mass bin, with a 0.7 correlation between the true and predicted mass.

We also apply the CNN and GBDT models to predict cluster optical depths. For this, we use a combination of features measured from tSZ, $\kappa$, and cluster virial mass. When including systematic uncertainties, in addition to the random realization of the detector noise and 10\% residual primary CMB, as simulated for the mass estimation, we also simulate a random realization of SO-like lensing convergence noise \citep{SimonsObservatory:2018koc} and 15\% scatter in the halo virial mass  \citep{Kravtsov:2014sra, Shi:2024bgm}. The uncertainty on $\tau_{pred}$ is, on average, 21$\%$ larger when we include systematic uncertainties. As shown in Fig.~\ref{fig:tau_M_pred}, the CNN and GBDT models provide optical depth predictions in agreement with the real values to well within the 1$\sigma$ variation in each optical depth bin both with and without systematic uncertainties included. The correlation between $\tau_{pred}$ and $\tau_{true}$ is 0.83 for both the CNN and GBDT models without systematic uncertainties and 0.78 for GBDT with systematic uncertainties included.
 
In Fig.~\ref{fig:tau_feat_importance}, we present how each feature helps the GBDT model to predict the optical depth. The prediction from tSZ signals alone (with no systematic noise) is used as the baseline, to calculate the ratio of the mean standard error (MSE) (\ref{eq:MSE}) between the baseline and the other scenarios using different sets of the cluster observable properties. We consider the model efficacy both with and without systematic errors for the different observables. By incorporating the deprojected CMB lensing convergence $\kappa$ map, the model's MSE decreases by $3\%$ on average relative to the baseline. The presence of white instrument noise and 10$\%$ CMB residuals in the tSZ signal degrades the optical depth estimation by 21$\%$ on average. The virial mass of the halo provides an extra valuable observable for optical depth estimation. Using all three observables in combination, the MSE is  $\sim$75\% ($\sim$60\%) of the baseline tSZ-only model with (without) systematic uncertainties. As a point of interest, we also show the idealistic optimal results one would get from the tSZ signal from the halo alone, without any other line of sight contributions, for three radii ($R_{vir}, R_{200}, R_{500}$). The errors are roughly $40\%$ of the baseline model. As shown in the right panel of Fig. \ref{fig:tau_feat_importance}, the fractional error on $\tau$ from the tSZ + $\kappa$ +$M_{vir}$ with (without) systematic noise is about $\sim$16\% ($\sim$14\%), while for the idealistic optimal scenario it is $\sim 10\%$.

We use the predicted optical depth estimates to infer the peculiar velocities for each halo using (\ref{eq:btauv}) 
\begin{equation}
\label{eq:btauv_pred}
  v_{pred} =\frac{c}{{\tau}_{pred}}\frac{\bar{b}_{AP(1.7')}}{A_{AP}}.
\end{equation}
Here $\tau_{pred}$ is the ML prediction for $\tau_{200}$. We assume $A_{ap} =2.37$ to correct for the signal attenuation in the aperture photometry and obtain an effective estimate of the disk kSZ signal, $\bar{b}_{200}\approx \bar{b}_{AP(1.7')}/A_{AP}$ so that $\bar{b}$ and $\tau_{pred}$ are consistently defined.  As discussed above, the approximation using a fixed aperture size for $\bar{b}$ introduces at most a 1\% error relative to using a varying aperture for the pairwise calculation. The predicted velocities are then used to calculate the predicted pairwise velocity, $V_{pred}$, as in (\ref{eq:Vhat}). 

\begin{figure}
\includegraphics[width = \columnwidth]{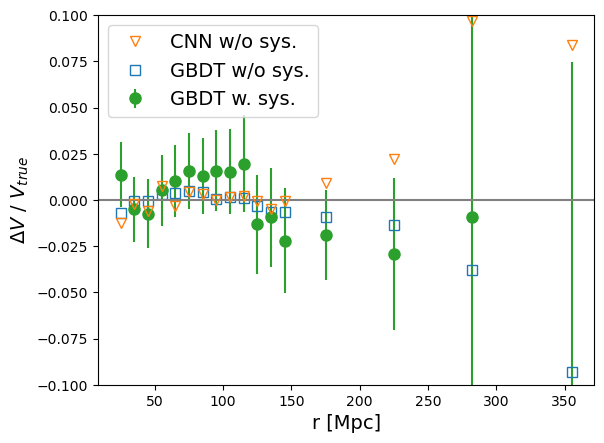}
\caption{The fractional difference, 
$\Delta{V}/V_{true} = 1- {V}_{pred}/{V}_{true} $, for the SO Test Sample between the estimator (${V}_{true}$) derived from the true velocity from the halo catalog and the prediction (${V}_{pred}$) derived from the machine learning model versus the truth for each cluster pair separation distance bin. Results are shown for the CNN and GBDT analyses without systematic contributions and for the GBDT with systematic uncertainties modeled for the tSZ, $\kappa$, $M_{vir}$ and $kSZ$ datasets.}
\label{fig:Vhat_kSZ}
\end{figure}

For the systematic error inclusion, following \citep{Gong:2023hse}, we create 10 independent noise realizations of primary CMB and white instrument noise to the tSZ and kSZ maps, along with lensing convergence noise and scatter associated in the virial masses. The 10 realizations are added to the clean maps and halo catalog data and the noisy features are used as inputs into the ML model. The predicted optical depths are then used to infer peculiar velocities for each halo for each realization. We calculate the mean of the pairwise velocity estimator averaged over the 10 realizations, and estimate errors by conducting a bootstrap on one realization.

Fig.~\ref{fig:Vhat_kSZ} compares the true pairwise velocity correlation, obtained directly from the velocities in the simulation, with those reconstructed from the kSZ maps and optical depths inferred from the GBDT and CNN baseline models with \{tSZ, lensing $\kappa$, virial mass\} both with and without systematic uncertainties included. The bootstrap errors on $V_{pred}$ are shown for the model with systematic errors. The GBDT model is able to recover an unbiased estimate of pairwise peculiar velocity within $2\%$ prediction uncertainty, which is well within the $1\sigma$ expected measurement uncertainties. The $\chi^2$ for the GBDT model is 10.9, i.e. a $\chi^2$ per degree of freedom of 0.64.  As a comparison we consider the $\chi^2$ for $V$ when estimated from the kSZ pairwise momentum and a single value of mass-averaged optical depth of the cluster sample inferred from the tSZ signal. For this, using the average optical depth of the cluster sample with an assumed $18\%$ uncertainty in optical depth estimates \citep{Vavagiakis:2021ilq, Flender:2016cjy}, the corresponding $\chi^2$ value is 28.9 ($\chi^2$ per degree of freedom 1.7). The result indicates that using the individual halo optical depths derived from the machine learning model provides a better fit to the true pairwise velocity than using a constant effective optical depth across the full sample and the approach remains robust in the presence of systematic uncertainties. 

\subsection{Application to ACT and SDSS data}
\label{sec:ACT}
We close by implementing the ML method to the ACT CMB maps and SDSS cluster catalogs analyzed in \citep{Calafut:2021wkx, Vavagiakis:2021ilq} to measure the pairwise velocity statistics. 

We use the set of cluster features measured with tSZ+$\kappa$ from ACT DR4 maps and $M_{vir}$ from SDSS DR15, to estimate the optical depth for each individual cluster using the GBDT models.  We also measure the kSZ temperatures using AP from the ACT DR5 CMB maps (which include ACT and Planck data). The tSZ and kSZ signals are isolated using AP, and the autoencoder is applied to the lensing convergence map to obtain the deprojected $\kappa$. We combine the ML-predicted cluster optical depth and kSZ temperatures, with the AP attenuation correction calibrated from the training simulation data in \ref{sec:training}, to estimate the peculiar velocities for each cluster as in (\ref{eq:btauv_pred}). 

\begin{figure}[!t]
\includegraphics[width = \columnwidth]{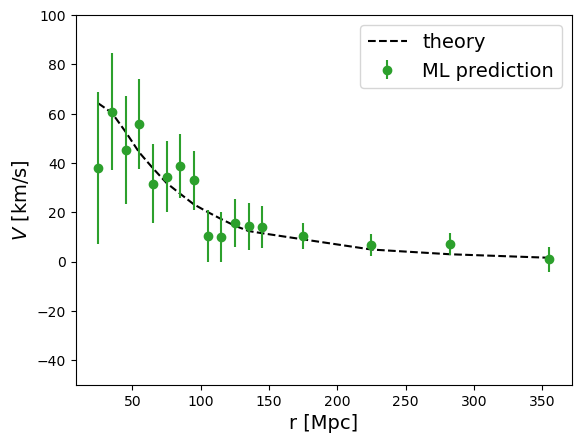}
\caption{The pairwise velocity statistics, and 1$\sigma$ uncertainties, for the ACT x SDSS data obtained from the ACT kSZ temperatures  and the GBDT-estimated optical depths, derived from ACT+SDSS [tSZ+$\kappa$+$M_{vir}$] data, for each cluster [green circles]. The theoretical prediction for the best fit Planck cosmology is also shown [black dashed line].
}
\label{fig:Vhat_ACT}
\end{figure}

Fig.~\ref{fig:Vhat_ACT}~ shows the pairwise velocity estimator for the sample for the ML-derived velocities. We find that the machine-learning model's estimate for the pairwise velocity statistic agrees well with the theoretical prediction derived from the linear theory using the Planck best-fit cosmology, with a $\chi^2$ relative to the theoretical prediction of 8.7 and SNR = 4.5.

\section{Conclusion}
\label{sec:conc}

In this work, we demonstrate the ability of machine learning modeling to estimate cluster mass and optical depth properties from 2D-projected  thermal SZ, kinematic SZ, and gravitational lensing data from the CMB and cluster catalog data from spectroscopic galaxy surveys. 

Aperture photometry (AP) is used to extract the SZ data. We find using the AP ring annulus directly connected to the central disk provides the highest SNR for the machine learning prediction.   An attenuation factor, to be applied to kSZ AP signal to account for the signal subtraction in the ring annulus relative to the unattenuated signal in the disk, is calibrated from the simulations and is well-modeled by a constant value across all cluster separation scales.

For CMB lensing, we develop an Autoencoder machine learning model to isolate the low redshift lensing convergence signal from the projected contamination of the large-scale structure from high redshifts. The deprojection increases the correlation between $\bar{\tau}_{200}$ and $\bar{\kappa}_{disk}$ by 68\%.

We investigate the inherent relations present in the simulated data prior to the machine learning analysis. The velocity-weighted optical depth $\bar{\tau}_{200}$ exhibits a lower correlation with the tSZ signal than $M_{200}$, which we attribute to the line of sight contributions unrelated to the cluster halo from the cylindrical integration. Strong correlations between the cylindrical quantities, $\bar{y}_{200}$ and $\bar{\tau}_{200}$, and the intrinsic cluster mass, $M_{200}$, exist however, indicating that cylindrical integration does not introduce significant measurement noise and the observables are viable features for the modeling. The map-derived observable, $\bar{y}_{AP}$, exhibits a positive correlation with $\bar{\tau}_{200}$ and $M_{200}$, however it is lower in magnitude than the idealized $y_{200}$ due to the measurement variance introduced by the AP filtering. 

The CNN and GBDT machine learning methods use different input features: CNN uses raw image cutouts from the sky maps, allowing the model to learn spatial patterns directly from the pixel intensities, while GBDT relies on averaged statistics measured sky map images. The mass prediction accuracy of the new CNN-ResNet architecture is  17$\%$ better than for the widely used CNN-VGG. Both the CNN-ResNet and GBDT models recover an unbiased cluster mass prediction for the test sample and have comparable accuracy for mass estimation. The CNN method is substantially more computational expensive than the GBDT, however, being roughly four orders of magnitude times faster. The GBDT model is shown to predict cluster mass with errors well within the estimated 1$\sigma$ measurement uncertainties (assuming a component-separated tSZ map with full detector noise and 10$\%$ residual primary CMB signals) down to $\sim 10^{13} M_{\odot}$.

The main focus of our work is to apply the ML methods to estimate the velocity-weighted optical depth for each cluster from the CMB and galaxy observables: the tSZ, the CMB lensing convergence $\kappa$ and halo virial mass estimates. We demonstrate the model efficacy and assess the prediction uncertainties for the optical depth with different combinations of these cluster properties as model features. The CNN and GBDT models provide unbiased estimates of the cluster optical depths. Using the GBDT model, the result is shown to be robust against the presence of systematic uncertainties in the model features, with prediction uncertainties smaller than the estimated  SO and DESI-modeled measurement uncertainties (assuming CMB detector noise and a 10$\%$ contribution from residual primary CMB signals for the tSZ signal, lensing convergence noise for $\kappa$, and 15$\%$ scatter in halo virial mass). With these systematic effects accounted for (or excluded), the three-observable model has MSE errors of under 75$\%$ (60\%) of the tSZ-only baseline model. 

We combine the optical depth estimates inferred from the GBDT model with the kSZ cluster temperatures estimated with AP to estimate the cluster peculiar velocities and the pairwise velocity correlation. The approach recovers an unbiased estimate of the pairwise velocity estimator with prediction errors well within the 1$\sigma$ measurement uncertainty (assuming the kSZ signal is extracted from maps with full primary CMB and detector noise). We compare the ML pairwise velocity correlation reconstruction against the standard method, in which the kSZ pairwise momentum is combined with a single, tSZ-inferred, mass-averaged optical depth value for the full cluster sample. We find that the machine learning prediction provides a significantly better match to the true pairwise correlation, with a lower $\chi^2$ value. With a $\chi^2$ per degree of freedom below 1, such that the ML prediction errors are smaller than the measurement uncertainties, it is demonstrated to be an effective method for measuring the pairwise velocity statistic from observational data.

We, finally, implement the approach using publicly available ACT, Planck and SDSS data and find that the model predicts a pairwise velocity correlation that is consistent with the theoretical statistic from the Planck best-fit cosmology. 

Our findings demonstrate that ML provides a promising and useful method to infer galaxy cluster properties, such as optical depth and mass, from astronomical observations.  Coincident high-fidelity simulations and synthetic data products, covering the range of objects and observables to be surveyed, are key to training and testing the algorithms.  With the wider, deeper, and more precise multi-wavelength observations of clusters expected with current and upcoming surveys (including DESI, Euclid, Roman, LSST, SO and CCAT), the opportunity exists to expand the range of ML features, including optical gravitational lensing data and X-ray data, among others. The pairwise velocity statistics that could be obtained in this way, across a range of redshifts, promise to provide valuable insights into the cosmic growth rate, with implications for understanding the nature of neutrinos models and gravity on cosmic scales. These ML tools can also be applied to a far broader set of statistics, to fully capitalize on the extraordinary astrophysical data promised in the coming decade.

\begin{acknowledgments}
We are grateful to Patricio Gallardo for providing the pairwise correlation code from \citep{Calafut:2021wkx} utilized in this work. The work of YG and RB is supported by NSF grant AST-2206088,  NASA grant 22-ROMAN11-0011, and NASA grant 12-EUCLID12-0004.
 
\end{acknowledgments}

\normalem
\bibliographystyle{apsrev}
\bibliography{ML}

\end{document}